\documentclass[aps,prb,twocolumn,superscriptaddress]{revtex4-2}
\usepackage{graphicx,epsfig}
\usepackage{amsmath, amssymb, amsfonts}
\usepackage[utf8]{inputenc}
\usepackage[T1]{fontenc}
\usepackage{comment}
\usepackage{titlesec}
\titleformat{\section}[hang]{\bfseries}{\thesection}{1em}{}

\newcommand {\be}{\begin{equation}}
\newcommand {\ee}{\end{equation}}
\newcommand {\bea}{\begin{array}}
\newcommand {\eea}{\end{array}}

\begin{document}

\title{ Interplay of Rashba and Dresselhaus Spin-Orbit Couplings on the Stability of Topological FFLO Phases in 1D Fermi Gases 
	}
\date{\today}

\author{Hamid Mosadeq}
\affiliation{Department of physics, Faculty of Science, Shahrekord university, Shahrekord 88186-34141, Iran }
\author{Mohammad-Hossein Zare}\email{zare@qut.ac.ir}
\affiliation{Department of Physics, Qom University of Technology, Qom 37181-46645, Iran}
\author{Reza Asgari}
\affiliation{Department of Physics, Zhejiang Normal University, Jinhua 321004, China}
\affiliation{School of Physics, Institute for Research in Fundamental Sciences (IPM), Tehran 19395-5531, Iran}

\begin{abstract}
	
We investigate the stabilization of topological Fulde-Ferrell-Larkin-Ovchinnikov (FFLO) phases, with a specific emphasis on the intraband FFLO phase, in a one-dimensional (1D) Fermi gas subjected to an external magnetic field. 
This research highlights the crucial role of the interplay between Rashba spin-orbit coupling (RSOC) and Dresselhaus spin-orbit coupling (DSOC).
Employing a Fermi-Hubbard model alongside the density matrix renormalization group (DMRG) method, we examine the combined effects of RSOC and DSOC on these exotic superfluid phases, taking into account attractive fermionic interactions. Our principal finding reveals that while RSOC primarily stabilizes conventional zero-momentum pairing, DSOC performs a distinct and crucial role in selectively stabilizing the intraband FFLO phase. This stabilization is achieved by enhancing spin polarization within a single helicity band and suppressing interband coherence, thereby facilitating the formation of finite-momentum FFLO pairs within the same band and resulting in the emergence of a topologically nontrivial superfluid. This targeted control of intraband FFLO pairing paves the way for new strategies in the manipulation of superfluid phases in spin-orbit coupled systems and offers essential insights for experimental realizations in ultracold atomic gases, with implications for topological quantum computing and Majorana fermions.
\end{abstract}


\maketitle

\section{introduction}

Majorana fermions are exotic quasiparticles~\cite{majorana1937teoria} that emerge in topological quantum systems~\cite{read2000paired,kitaev2001unpaired,wilczek2009majorana} and are important candidates for fault-tolerant topological quantum computation~\cite{nayak2008non}. While they were originally proposed as elementary particles, they now appear as collective excitations in systems like topological superconductors and fractional quantum Hall states. Their relevance to this study lies in the role they may play in systems with spin-orbit coupling and FFLO phases.

Several theoretical proposals have predicted the emergence of non-Abelian Majorana zero modes at interfaces of hybrid structures involving conventional s-wave superconductors, topological insulators, semiconductors, and ferromagnetic materials~\cite{fu2008superconducting,fu2009josephson,cook2011majorana,sun2016majorana,lutchyn2010majorana,oreg2010helical,biderang2018edge}.
These zero modes arise via the proximity effect.
Experimental studies have reported zero-bias conductance peaks in InSb nanowires coupled to superconducting leads- signatures consistent with Majorana fermions~\cite{Mourik2012,rokhinson2012fractional,deng2012anomalous,churchill2013superconductor,das2012zero,zhang2017ballistic}.
Meanwhile, ultracold atomic gases, known for their tunability and isolation from the environment, offer promising platforms for realizing and controlling Majorana modes~\cite{jiang2011majorana}.
To realize topological phases in such systems, two ingredients are essential~\cite{oreg2010helical,lutchyn2010majorana}: (i) a spin-orbit coupled single-particle dispersion in the presence of a Zeeman field, (ii) pairing correlations, typically induced via proximity to an s-wave superconductor.
Ultracold atomic systems naturally host attractive interactions, which can be tuned via Feshbach resonances~\cite{fu2014production,chin2010feshbach}, and synthetic spin-orbit coupling (SOC) and Zeeman fields can be engineered using laser setups in 1D geometries~\cite{wang2012spin,cheuk2012spin,cui2013synthetic,zhai2015degenerate}. 

The successful implementation of SOC in ultracold atomic gases has paved the way for exploring of exotic superfluid phases~\cite{galitski2013spin}, including topological Bardeen-Cooper-Schrieffer (BCS) states and FFLO superfluids~\cite{fulde1964superconductivity,larkin1964nonuniform}, within highly controllable experimental platforms~\cite{gong2011bcs,hu2011probing,yu2011spin,gong2012searching,vyasanakere2011bcs,qu2013topological}.
In BCS theory, fermions form Cooper pairs with opposite spin and momentum. However, the application of a magnetic field introduces spin imbalance, which shift the Fermi surfaces and can disrupt conventional pairing.
 Under certain conditions, this imbalance leads to the formation of FFLO states- inhomogeneous superfluid phases where Cooper pairs acquire finite center-of-mass momentum. Such phases are particularly favored in 1D systems with attractive interactions, which show a rich phase diagram.
   As the Zeeman field increases, the system transitions from a conventional BCS phase to a partially polarized phase and, eventually, to a fully polarized phase~\cite{esslinger2010fermi,liao2010spin,feiguin2012bcs}. The intermediate phase may host an FFLO state, characterized by finite-momentum pairing and distinct twin peaks in the pair momentum distribution.
   Moreover, topologically nontrivial superfluid phases can emerge in 1D systems with RSOC~\cite{zhang2008px,sato2009non,chen2013inhomogeneous}, especially when the chemical potential is tuned to occupy only a single helicity band, effectively inducing triplet p-wave pairing.
The FFLO phase is further stabilized in the presence of on-site attractive interactions, such as those modeled by the attractive 1D Hubbard model~\cite{liang2015unconventional,hu2015phase,chan2015numerical,singh2024fulde}.

 While the role of RSOC in stabilizing topological FFLO phases has been studied extensively, the combined influence of RSOC and DSOC— particularly their competition— remains poorly understood. 
 Motivated by recent experiments on synthetic SOC with tunable RSOC and DSOC in cold atom systems~\cite{white1992density,schollwock2005density}, we investigate how this interplay affects the stability of FFLO phases in a 1D attractive Fermi gas subject to a Zeeman field. We consider a Fermi-Hubbard model incorporating both types of SOC and employ the DMRG method based on matrix product states to study the emergent phases. This method, well-suited for strongly correlated 1D systems, allows us to map out the phase diagram with high numerical precision. We perform simulations on open chains up to length $L=300$, with bond dimensions up to $M=400$, and carry out 20 sweeps to ensure convergence.

 We identify several distinct pairing phases, including conventional BCS superfluid, intraband FFLO, and \emph{mixed-FFLO} phases, and analyze how the interplay between RSOC and DSOC affects their stability. The \emph{mixed-FFLO} phase is characterized by the coexistence of both intraband and interband FFLO pairings, reflecting the complex interplay between RSOC and DSOC. 
 Our results show that the interplay between RSOC and DSOC governs the stability of various pairing states. While RSOC stabilizes zero-momentum pairing, DSOC favors intraband FFLO pairing by enhancing spin polarization and suppressing interband coherence.
  This work provides a detailed characterization of the interplay between RSOC and DSOC in 1D quantum systems. It also offers experimentally relevant guidance for realizing tunable superfluid phases in spin-orbit coupled ultracold gases.
 
 This paper is structured as follows: Section II introduces the model Hamiltonian, providing the foundation for the study. Section III then presents the phase diagrams, analyzing the characteristics of the superfluid phases that emerge from the model."
  Additionally, we discuss the topological properties of the system and the signatures of Majorana modes. Finally, Section IV concludes the paper with a summary of our findings and discusses their implications for future research and experimental realizations.

\section{MODEL and Physical Quantities}
\label{sec2}
\subsection{Model}
We consider a 1D Fermi gas of ultracold atoms confined in an optical lattice subjected to an external magnetic field. The system is described by a Fermi-Hubbard Hamiltonian that incorporates Zeeman field $h$, both RSOC and DSOC, and on-site interaction terms. The Hamiltonian is given by~\cite{chen2013inhomogeneous,alicea2010majorana,sato2009non,alicea2012new,wei2012majorana,liu2012topological,wu2013unconventional}
\begin{equation}
	H=H_{\rm K}+H_{\rm R}+H_{\rm D}+H_{\rm U}
	\label{eq:ham}
\end{equation}
where the individual terms are:

(i) Kinetic Energy Term $H_{\rm K}$:
\begin{equation}
		 H_K=-t\sum_{i,j,\sigma} c_{i\sigma}^\dag c_{j\sigma}+\sum_{i,\sigma}(h\sigma_z-\mu)c_{i\sigma}^\dag c_{i\sigma}
		\label{eq:hamd1}
\end{equation}
here, $c_{i\sigma}$ and $c_{i\sigma}^\dag$ are fermionic annihilation and creation operators, respectively, for spin $\sigma=\uparrow,\downarrow$ at site $i$.
 The term $n_{i,\sigma}=c_{i\sigma}^\dag c_{i\sigma}$ is the number operator. 
The parameter $t$ represents the hopping amplitude between nearest neighbors, $h$ is the Zeeman field $h$, and $\mu$ is the chemical potential.  

(ii) Rashba Spin-Orbit Coupling Term $H_{\rm R}$: 

\begin{equation}
			H_{\rm R}=-\alpha\sum_i(c_{i\downarrow}^\dag c_{i+1\uparrow}-c_{i\uparrow}^\dag c_{i+1\downarrow}+ {\rm H.c.})
			\label{eq:hamd3}
\end{equation}
this term describes the spin-flip RSOC~\cite{bychkov1984oscillatory} with strength $\alpha$, which introduces spin-dependent hopping.

(iii) Dresselhaous Spin-Orbit Coupling Term $H_{\rm D}$:

\begin{equation}
				H_{\rm D}=-\beta\sum_i( ic_{i\uparrow}^\dag c_{i+1\uparrow}- ic_{i\downarrow}^\dag c_{i+1\downarrow}+{\rm H.c.})
			\label{eq:hamd3}
\end{equation}
this term accounts for the spin-conserved Dresselhaus (110)~\cite{dresselhaus1955spin}, with strength $\beta$, which couples spin states differently compared to RSOC and introduces an additional spin-momentum locking effect.

(iv) On-Site Interaction Term $H_{\rm U}$: 

\begin{equation}
		H_U=U\sum_i n_{i\uparrow}n_{i\downarrow} 
	\label{eq:hamd4}
\end{equation}
this term represents the on-site Coulomb interaction between spin-up and spin-down fermions, where $U$ is the interaction strength.

The total Hamiltonian described above models a 1D spin-imbalanced Fermi gas with both Rashba and Dresselhaus spin-orbit couplings (SOCs), under the influence of a Zeeman field. These interactions, along with the attractive on-site interaction, lead to the formation of superfluid phases, with particular emphasis on the FFLO phase.
Throughout this work, the hopping integral $t$ is
set to unity as the energy scale.

\subsection{Key Physical Quantities}

This study investigates the quantum phase diagrams of the Fermi-Hubbard model for 1D spin-1/2 Fermi gases with attractive interparticle interactions. At low temperatures, these gases exhibit superfluid properties, where fermions pair up. These pairs typically form between spin-up and spin-down fermions, and in systems with balanced spin populations, this process is well explained by conventional BCS theory. However, when there is an imbalance in the number of spin-up and spin-down fermions (such as spin polarization), the pairing mechanism changes.

\emph{Spin Polarization}:
Spin polarization is a crucial factor in understanding these systems. The spin polarization, denoted by the parameter $p$, is defined as:

\begin{equation}
	p = \frac{N_{\uparrow} - N_{\downarrow}}{N_{\uparrow} + N_{\downarrow}}
\end{equation}
where $N_{\uparrow}$ and $N_{\downarrow}$ are the numbers of spin-up and spin-down particles, respectively. Superfluidity in these systems is expected to collapse if the polarization exceeds a critical threshold~\cite{radzihovsky2010imbalanced,chevy2010ultra,gubbels2013imbalanced}. Below this threshold, polarized superfluid states may emerge. One particularly interesting state is the FFLO phase, in which pairs of fermions possess a non-zero center-of-mass momentum $k_{FFLO}$, leading to a spatially modulated order parameter~\cite{roscher2015phase}. The FFLO phase has been extensively studied and confirmed in both theoretical~\cite{rizzi2008fulde,tezuka2008density,tezuka2010ground,batrouni2008exact,casula2008quantum,heidrich2010phase} and experimental~\cite{guan2007phase,kakashvili2009paired,lee2011asymptotic,cheng2018fulde} work for spin-imbalanced systems.

\emph{Pairing and Momentum Distribution:}
In 1D systems, long-range fluctuations suppress spontaneous symmetry breaking~\cite{mermin1966absence,hohenberg1967existence}, making it challenging to characterize the pairing state using traditional order parameters. Therefore, we analyze pairing correlations both in real space and momentum space.

\subsubsection{Real-Space Pairing Correlations}
These correlations reveal the spatial structure of fermion pairs and are defined as:

\begin{equation}
P^{\text{pair}}_{ll'} = \langle \Delta_l^{\dagger} \Delta_{l'} \rangle	
\end{equation}

\subsubsection{Momentum-Space Pairing Correlations}
The Fourier transform of the real-space pairing correlations gives us the pair momentum distribution (PMD)~\cite{feiguin2007pairing,rizzi2008fulde,tezuka2008density,tezuka2010ground}:

\begin{equation}
n^{\text{pair}}_k = \frac{1}{L} \sum_{l, l'} e^{ik(l-l')} P^{\text{pair}}_{ll'}
\end{equation}

This distribution serves as a powerful tool to distinguish between different types of pairing states:

\begin{itemize}
	\item A single peak at $k = 0$ indicates conventional BCS pairing, where the fermion pairs have zero momentum.
	\item Two peaks at finite momenta indicate unconventional pairing, such as the FFLO state, where the pairs have a finite center-of-mass momentum $\pm k_{FFLO}$.
	\item In some cases, we may observe a mixed pairing (MP) phase, where both BCS and FFLO pairing phase coexist. This phase is characterized by three distinct peaks: one at $k = 0$ (BCS-type pairing) and two at $±k_{FFLO}$ (FFLO-type pairing).
\end{itemize}

\emph{Thermodynamic Quantities:}
To study the phase transitions between distinct phases, we calculate thermodynamic quantities derived from the energy density:

\begin{itemize}
	\item Particle Density $n$: The particle density is given by the first derivative of energy with respect to chemical potential:
	
	\begin{equation}
			n = -\frac{\partial E}{\partial \mu}
	\end{equation}
	
	\item Compressibility $\chi_{\mu}$: Compressibility is the second derivative of energy with respect to chemical potential:
	
    \begin{equation}
    		\chi_{\mu} = -\frac{\partial^2 E}{\partial \mu^2}
    \end{equation}

\end{itemize}

These quantities help map out the phases of the system and identify the stability regions for various phases such as BCS, FFLO, and the normal gas phase.

\emph{Entanglement Spectrum:}
The topological properties of the superfluid phase are investigated by computing the entanglement spectrum from the many-body wave function. The entanglement spectrum provides information about the quantum correlations in the system and can reveal signatures of topologically nontrivial phases. In topological phases, the entanglement spectrum exhibits a twofold degeneracy~\cite{turner2011topological}, which is a precursor to the emergence of zero-energy Majorana edge states~\cite{qi2012general}.
The entanglement spectrum is subsequently derived from the
eigenvalues of the entanglement Hamiltonian, $H_{\rm E}=-\ln {\rho_{L}}$.
The reduced density matrix of the left subsystem, $\rho_{L}$, is defined as ${\rm Tr}_{R}[|\psi\rangle \langle \psi|]$, where the trace is taken over all sites in the right half of the 1D system.

\section{ Results}

\subsection{Band Structure and Pairing Mechanisms}

We present a comprehensive analysis of the phase diagrams of 1D Fermi gases, derived using the numerical DMRG method applied to the Fermi-Hubbard Hamiltonian (\ref{eq:ham}).
The phase diagram is constructed by considering fillings \(n \leq 1\). Notably, the particle-hole symmetry of the model, which remains invariant under the transformations \(c_{i,\sigma} \to \sigma(-1)^i c_{i, -\sigma}^\dagger\), allows us to extend the phase diagram for fillings above half-filling by applying a particle-hole transformation.

To explore potential exotic pairing mechanisms, we examine the single-particle spectrum:
\begin{equation}
	\epsilon(k) = -2t \cos k \pm \sqrt{(2\alpha \sin k)^2 + (h - 2\beta \sin k)^2},
\end{equation}
which is derived from the non-interacting Hamiltonian in momentum space:
\begin{equation}
H_0(k) = -2t \cos k \sigma_I - 2\alpha \sin k \sigma_y + (h - 2\beta \sin k) \sigma_z.	
\label{eq:kham}
\end{equation}

In the absence of SOCs, the application of a magnetic field causes the spectrum to split into two fully spin-polarized bands along the \( \pm \sigma_z \) directions.
When considering the attractive interaction, it becomes apparent that only interband finite-momentum pairings are feasible.  RSOC and DSOC, represented by the second and third terms on the right-hand side of (\ref{eq:kham}), respectively, effectively function as momentum-dependent magnetic fields. These fields have magnitudes of \( -2\alpha \sin k \) for RSOC and \( -2\beta \sin k \) for DSOC, coupling to the spin along the \(y\) and \(z\) directions, respectively.

\begin{figure}[]
	\centering
	\includegraphics[scale=.9]{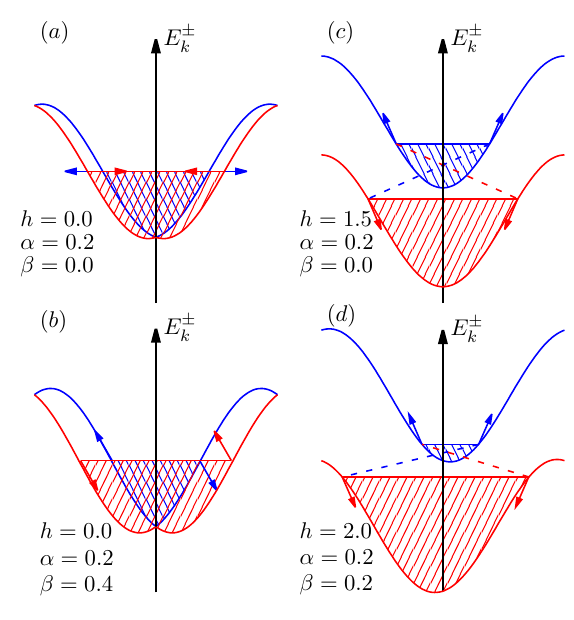}
	\caption{ (Color online)
		Single-particle band structures illustrating unconventional pairings in a 1D Fermi gas of ultracold atoms. The panels depict band structures corresponding to various parameter sets.
		%
        Arrows represent the spin polarization vectors (or spin orientation) specifically evaluated at the Fermi points.		    }
	\label{fig:rd_spec}
\end{figure}

As a result, this effective magnetic field induces momentum-dependent spin polarizations:

\begin{align}
	{\bf S}_{\pm} = & \pm \frac{1}{\sqrt{4\alpha^2 \sin^{2}k+(h-2\beta\sin k)^2}} 
	\nonumber  \\ 
	& \times (0,2\alpha\sin k, h-2\beta\sin k )
\end{align}
where the \( \pm \) signs correspond to distinct bands. Since no spin polarization occurs along the \(x\)-direction, the spin-polarization angles at the Fermi points are defined as:

\begin{equation}
\theta_{\pm} = \tan^{-1}\left(\frac{2\alpha \sin k}{h - 2\beta \sin k}\right). 
\end{equation}

These angles, \( \theta_{\pm} \), represent the spin-polarization directions at the Fermi points relative to the \(z\)-axis. The physics becomes particularly intriguing when both bands are partially occupied, as the interplay between RSOC and DSOC leads to distinct spin-polarization effects that modulate pairing mechanisms.

To accurately characterize the intrinsic pairings within Hamiltonian (\ref{eq:ham}), these spin-polarizations are crucial. Fig.~\ref{fig:rd_spec} presents the single-particle band structures derived from the non-interacting Hamiltonian (\ref{eq:kham}) at half-filling. Under these conditions, both bands exhibit partial occupancy, leading to the formation of two distinct Fermi surfaces, denoted by \( \epsilon_{F \pm} \). These surfaces give rise to four distinct Fermi points: \( \pm k_1 \) and \( \pm k_2 \).

When considering only RSOC (or DSOC), the band structure splits and the spin configuration aligns either in-plane (RSOC) or out-of-plane (DSOC), as shown in Fig.~\ref{fig:rd_spec}(a) (the case of DSOC is not explicitly depicted). In both cases, the introduction of an attractive interaction facilitates the formation of intraband BCS (\emph{intra}-BCS) pairing. However, when both RSOC and DSOC are present, the spin orientation shifts from an in-plane configuration to one aligned in the \(xz\)-plane, as depicted in Fig.~\ref{fig:rd_spec}(b). Importantly, \emph{intra}-BCS pairing remains stable even in the presence of significant DSOC.

Fig.~\ref{fig:rd_spec}(c) illustrates the single-particle band structure that includes both RSOC and a Zeeman field. In this configuration, the spins at the Fermi points align within the \(xz\)-plane, enabling the formation of both interband FFLO (\emph{inter}-FFLO) pairings and \emph{intra}-BCS pairings. The coexistence of RSOC and a magnetic field results in the opening of a gap at \(k = 0\), potentially facilitating a phase transition between the \emph{intra}-BCS and \emph{inter}-FFLO pairing states, which can be tuned by adjusting the strength of RSOC or the magnetic field. This analysis shows that the coexistence of \emph{intra}-BCS and \emph{inter}-FFLO pairings may be feasible, depending on the filling factor.

Similarly, the inclusion of DSOC and a magnetic field also opens a gap at \(k = 0\) (not shown), thereby promoting the stabilization of the \emph{inter}-FFLO phase. Adjusting the strength of DSOC or the magnetic field further tunes the system towards a stable \emph{inter}-FFLO state.

In Fig.~\ref{fig:rd_spec}(d), we show the single-particle band structure with both RSOC, DSOC, and a Zeeman field. Our analysis reveals that even small DSOC induces a symmetry-breaking in the energy bands at \(k = 0\), which plays a crucial role in facilitating a phase transition from  \emph{intra}-BCS pairing to \emph{intra}-FFLO pairing. This leads to the formation of the \emph{mixed}-FFLO phase, where both \emph{intra}- and \emph{inter}-FFLO pairings coexist.

\begin{figure}[]
	\centering.
	\includegraphics[scale=.9]{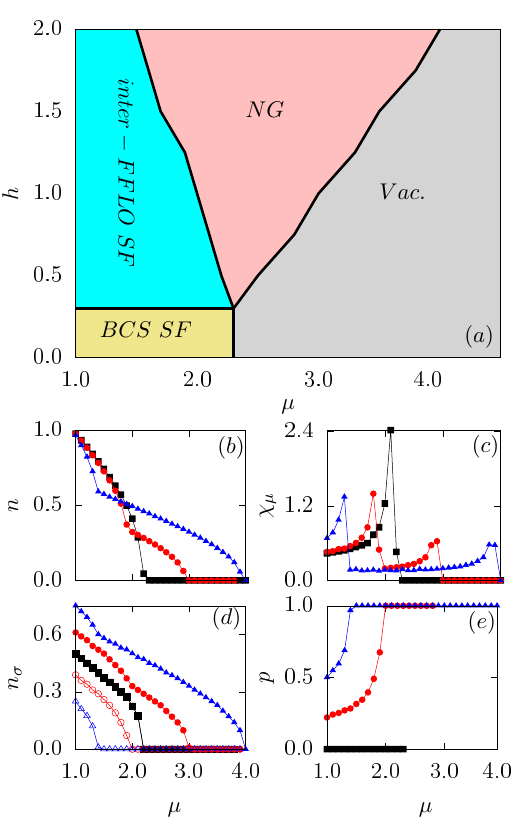}
	\caption{ (Color online) (a)
		     (a) Phase diagram of a 1D Fermi gas system in the $h-\mu$ plane, for $U = -2$ and zero SOCs. Panels (b)-(e) illustrate the properties of a chain with $L = 40$ sites as a function of the chemical potential $\mu$: (b) fermion population $n$, (c) compressibility $\chi_{\mu}$, (d) spin densities ($n_{\uparrow},n_{\downarrow}$), and (e) spin polarization $p$ for various magnetic field strengths: $h = 0.0$ (squares), $h = 1.0$ (circles), and $h = 2.0$ (triangles). In panel (d), $n_{\uparrow}$ is represented by open symbols, while $n_{\downarrow}$ is represented by filled symbols.}
	\label{fig:pt1}
\end{figure}

\begin{figure}[]
	\centering.
	\includegraphics[scale=.77]{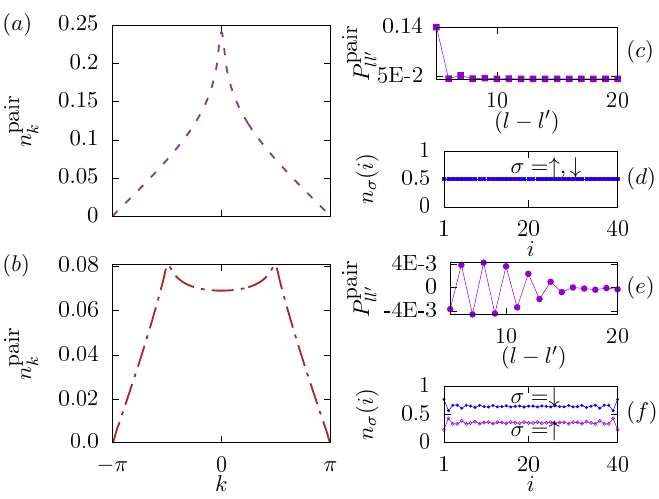}
	\caption{ (Color online) 
		(a,b) Pair momentum distribution, $n_{k}^{\rm pair}$: The PMD in momentum space is presented for (a) the BCS state $(h=0, \mu=1)$ and (b) the \emph{inter}-FFLO phase $(h=2, \mu=1)$, corresponding to specific points in the main phase diagram illustrated in [Fig.~\ref{fig:pt1}(a)].
		(c-f) Real-space correlations ($P_{ll'}^{\rm pair}$) and spin densities ($n_{\uparrow},n_{\downarrow}$) for an $L=40$ site chain: These panels depict the real-space properties for selected points from the BCS and \emph{inter}-FFLO phase regions of the main phase diagram shown in [Fig.~\ref{fig:pt1}(a)]. Specifically, panels (c) and (e) illustrate the real-space pairing correlations, while panels (d) and (f) present the corresponding spin densities. Panels (c) and (d) characterize the BCS phase, whereas panels (e) and (f) represent the \emph{inter}-FFLO phase.
		} 
	\label{fig:pt2}
\end{figure}

\subsection{Phase Diagrams of 1D Fermi Gases in the Absence of Spin-Orbit Coupling}

Fig.~\ref{fig:pt1}(a) illustrates the phase diagram of 1D Fermi gases in the absence of SOC, where the SOC parameters \(\alpha\) and \(\beta\) are both set to zero, for a fixed interaction strength of \(U = -2\). In the \(h-\mu\) plane, four distinct phases are observed: (i) the conventional BCS superfluid, (ii) the \emph{inter}-FFLO superfluid, (iii) the vacuum phase (empty bands without filling), and (iv) the normal gas (NG) phase.

In the absence of an external magnetic field, the system remains in the conventional BCS phase, characterized by the formation of Cooper pairs with zero center-of-mass momentum. This phase remains stable for chemical potentials \(\mu \leq 2.3\). Within this parameter range, the PMD, \(n^{\text{pair}}_k\), displays a single peak at \(k = 0\), which is a hallmark of the conventional BCS phase, as illustrated in Fig.~\ref{fig:pt2}(a). As the magnetic field \(h\) is gradually increased, the profile of \(n^{\text{pair}}_k\) evolves into a dual-peaked structure, with maxima at finite momenta. This transformation signifies the emergence of the \emph{inter}-FFLO phase, as shown in Fig.~\ref{fig:pt2}(b).

To better understand the characteristics of these two phases, we examine the real-space pairing correlation function \(P^{\text{pair}}_{ll'}\) and the spatial distribution of spin densities \(n_{\sigma}(i)\) for fixed values of \(\mu\) and \(h\) that correspond to the BCS and \emph{inter}-FFLO phases. In the BCS phase, the pairing correlation function exhibits no oscillatory behavior and decays according to a power law with respect to \(|l - l'|\), with no discernible nodal points, as shown in Fig.~\ref{fig:pt2}(c). Additionally, no spin imbalance is observed in the system, as depicted in Fig.~\ref{fig:pt2}(d).

In contrast, the real-space pairing correlation function associated with the \emph{inter}-FFLO phase displays pronounced oscillations in both magnitude and sign, as shown in Fig.~\ref{fig:pt2}(e). Moreover, the spin densities for up- and down-spin components are distinctly separated, confirming that the system resides in a nontrivial \emph{inter}-FFLO phase, which is a magnetic state, as shown in Fig.~\ref{fig:pt2}(f). 
The formation of unconventional pairing in moderate magnetic fields is attributed to the Zeeman-induced splitting of energy bands, which results in the emergence of two fully spin-polarized bands along the \(z\)-direction. Consequently, pairing is restricted to fermions originating from distinct bands, facilitated by inter-species interactions, thus leading to the well-established \emph{inter}-FFLO phase.

Fig.~\ref{fig:pt1}(b) demonstrates the evolution of the fermion population \(n\) for chemical potentials exceeding half-filling (\(\mu > 1\)). As the chemical potential increases, the fermion population monotonically decreases, ultimately reaching zero at \(\mu = 2.3\) for a constant magnetic field \(h = 0\). For \(\mu \geq 2.3\), the system transitions into the vacuum phase, which is denoted as "Vac." in the phase diagram.
 This finding is corroborated by the anomaly in compressibility \(\chi_\mu\), as shown in Fig.~\ref{fig:pt1}(c), which aligns precisely with the phase transition point from the BCS phase to the vacuum phase in Fig.~\ref{fig:pt1}(a). Furthermore, Fig.~\ref{fig:pt1}(d) shows that for \(\mu < 2.3\), the observed equality between up-spin and down-spin densities substantiates the BCS state as the plausible ground state for a spin-balanced system with nonzero attractive on-site interactions.

Fig.~\ref{fig:pt1}(c) also demonstrates the presence of two distinct singularities in the compressibility \(\chi_\mu\) for \(h = 1\), which correspond to phase transitions in the phase diagram shown in Fig.~\ref{fig:pt1}(a). This evidence, together with the observed disparity between up-spin and down-spin densities for \(h = 1\) as illustrated in Fig.~\ref{fig:pt1}(d), provides compelling support for the \emph{inter}-FFLO state as a feasible ground state in a spin-imbalanced system characterized by nonzero attractive on-site interactions. Within the \emph{inter}-FFLO phase, the polarization gradually increases and saturates at \(\mu = 1.8\), as shown in Fig.~\ref{fig:pt1}(e). This transition leads to the NG state, characterized by a vanishing charge gap for \(U < 0\) (not shown). The behavior of polarization is consistent with theoretical predictions for both the \emph{inter}-FFLO and NG phases. Notably, the peak of the PMD at finite momenta diminishes and eventually vanishes at the phase boundary separating the \emph{inter}-FFLO phase from the NG phase. Consequently, Cooper pairs are absent in the NG phase. As the chemical potential increases, a phase transition occurs from the NG state to the vacuum phase at a critical value of \(\mu\), where the fermion population \(n\) approaches zero, as illustrated in Fig.~\ref{fig:pt1}(b).

The stability region of the intermediate NG phase expands with increasing \(h\), due to the enhanced polarization strength. It is noteworthy that our results indicate a reduction in the stability region of the \emph{inter}-FFLO phase with an increase in attractive on-site Hubbard interaction. For instance, the phase boundary between the BCS phase and the \emph{inter}-FFLO phase shifts to \(h = 0.7\) for \(U = -4\). We further investigate the finite-size effects on the phase boundaries shown in Fig.~\ref{fig:pt1}(a), finding that as the system size increases, the compressibility peaks become progressively narrower and sharper, with subtle shifts in their positions.

\begin{figure}[]
	\centering.
	\includegraphics[scale=.52]{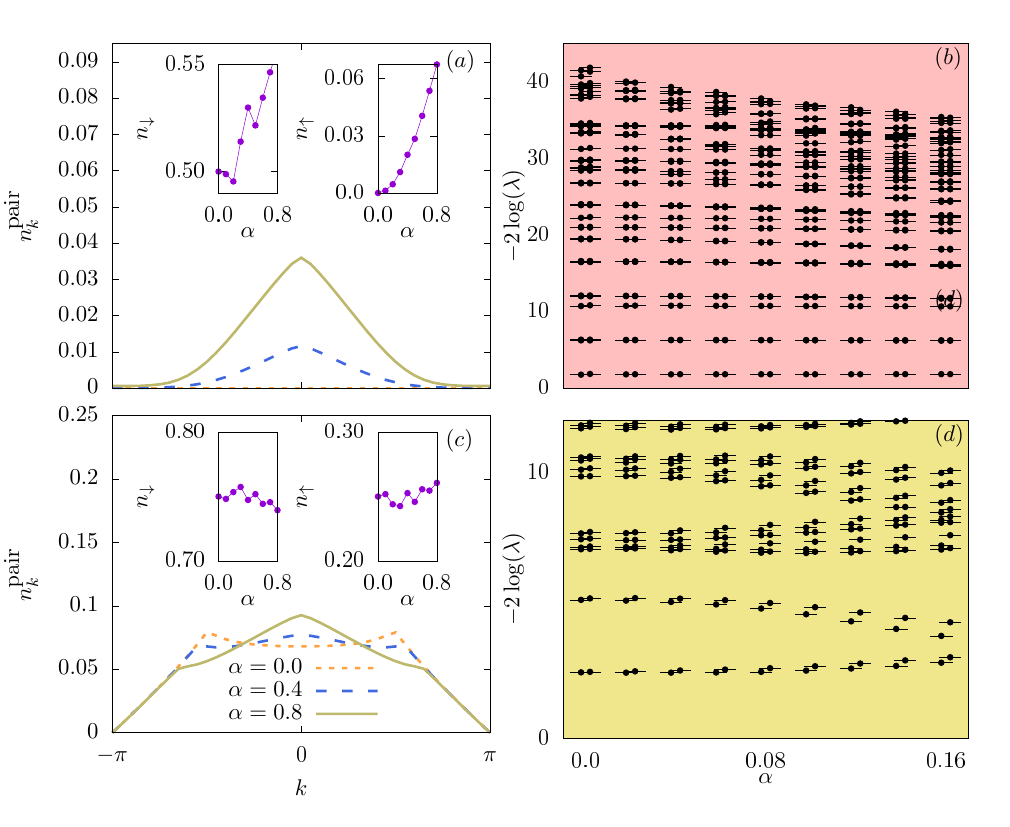}
	\vspace{-0.5cm}
	\caption{ (Color online) 
	(a,c) Pair momentum distribution, $n_{k}^{\text{pair}}$: Evaluation of the PMD in momentum space for specific regions of the main phase diagram [Fig.~\ref{fig:pt1}(a)] while varying the RSOC strengths: $\alpha= 0$ (dotted line), $\alpha=0.4$ (dashed line), and $\alpha= 0.8$ (solid line).
	Panel (a) pertains to the NG region ($h=2.0$, $\mu=2.0$), whereas (c) illustrates the \emph{inter}-FFLO region ($h=2.0$, $\mu=1.0$).
	(b,d) Entanglement spectrum, $-2\log(\lambda)$: The entanglement spectrum as a function of the RSOC strength ($\alpha$), corresponding to the same phase diagram points as panels (a) and (c).
	 Panel (b) represents the NG region ($h=2.0$ and $\mu=2.0$), while panel (d) depicts the \emph{inter}-FFLO region ($h=2.0$ and $\mu=1.0$).
	 The dots in panels (b,d) denote the eigenvalues of the entanglement Hamilatonian, $H_{\rm E}=-\ln \rho_{\rm L}$. In topological phases, the entanglement spectrum is expected to exhibit a twofold degeneracy. 
	Insets: Each panel is accompanied by an inset that illustrates the spin densities $(n_{\uparrow},n_{\downarrow})$ as a function of the RSOC strength.}
	\label{fig:plz}
\end{figure}

\subsection{Effect of RSOC and DSOC on the Phase Diagram of 1D Fermi Gases}

In this section, we examine the distinct roles of RSOC and DSOC on the phase diagram of a 1D Fermi gas [Fig.~\ref{fig:pt1}(a)]. Starting with a fixed interaction strength of \( U = -2 \) and setting \( \beta = 0 \) for DSOC, we investigate the impact of RSOC on the system's behavior. Previous studies on half-filled 1D lattice systems have demonstrated that RSOC contributes to the stabilization of the BCS pairing phase~\cite{liang2015unconventional}. 
For weak and moderate RSOC strengths, both \emph{inter}-FFLO and \emph{intra}-BCS pairings can coexist, resulting in a mixed pairing state. However, as the RSOC strength increases substantially, BCS pairing tends to dominate.

To understand the effect of RSOC on the phase diagram [Fig.~\ref{fig:pt1}(a)], we calculate the pair momentum distribution across a range of chemical potential (\( \mu \)) and Zeeman field (\( h \)) values. These results, shown in Fig.~\ref{fig:plz}, provide insights into various phases within the diagram.
\begin{figure}[]
	\centering
	\includegraphics[scale=0.9]{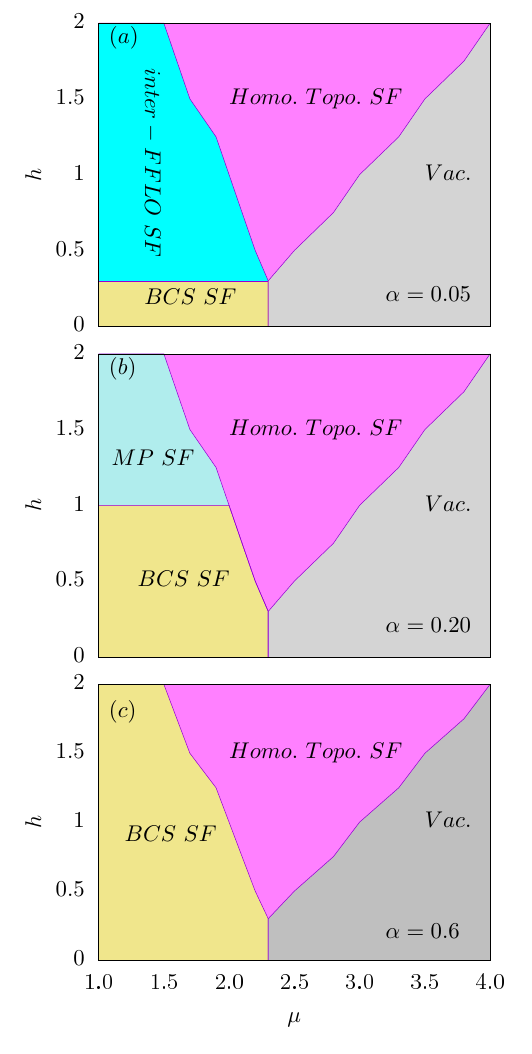}
	\caption{ (Color online) Phase diagrams of a 1D Fermi gas system in the $h-\mu$ plane. These diagrams are constructed for fixed parameters $U = -2$ and $\beta=0$, while varying the strengths of RSOC: (a) $\alpha=0.05$, (b) $\alpha=0.2$, and (c) $\alpha=0.6$.}
	\label{fig:pt4}
\end{figure}
\begin{figure}
	\centering
	\includegraphics[scale=.75]{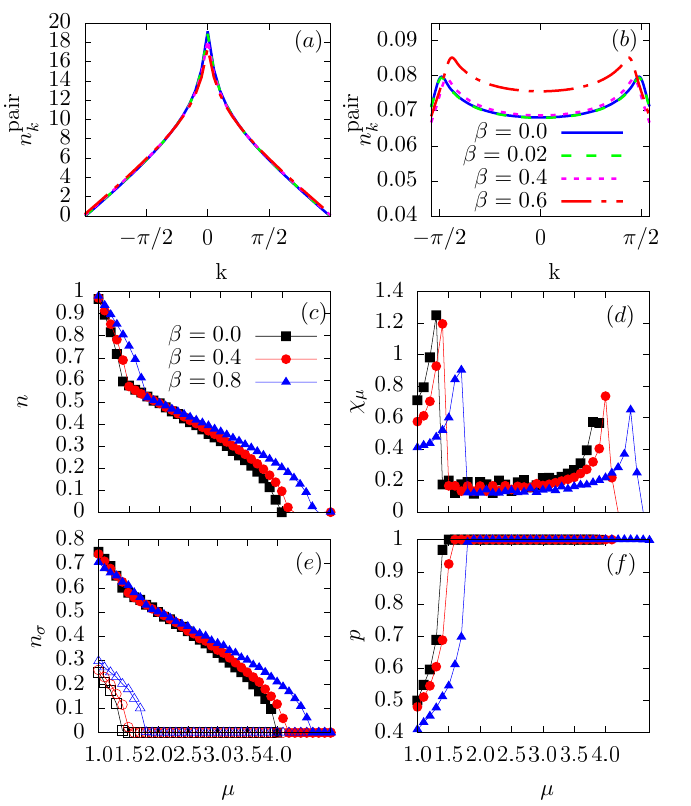}
	\caption{ (Color online)
		(a-b) Pair momentum distribution, $n_{k}^{\text{pair}}$: Evaluation of the PMD in momentum space for specific regions of the main phase diagram [Fig.~\ref{fig:pt1}(a)] across varying the DSOC strength: $\beta= 0$ (solid line), $\beta=0.2$ (dashed line), $\beta=0.4$ (dotted line), and $\beta= 0.6$ (dash-dotted line).
		Panel (a) corresponds to the BCS region ($h=0.0$, $\mu=1.0$), while panel (b) depicts the \emph{inter}-FFLO region ($h=2.0$, $\mu=1.0$).
		(c-f) Dependence on chemical potential $\mu$: These panes illustrate the dependence of key properties on $\mu$ for a chain with $L=40$ with $U = -2$ and $\alpha = 0$ at magnetic field $h = 2.0$. The following properties are presented: (c) fermion population $n$, (d) compressibility $\chi_{\mu}$, (e) spin densities ($n_{\uparrow},~ n_{\downarrow}$), and (f) spin polarization $p$.
		Various strengths of DSOC are represented by different symbols: $\beta = 0.0$ (squares), $\beta = 0.4$ (circles), and $\beta = 0.8$ (triangles). In panel (f), open symbols represent $n_{\uparrow}$ while filled symbols denote $n_{\downarrow}$.}
	\label{fig:dsoc}
\end{figure}
For the specific parameter values \( h = 2 \) and \( \mu = 2 \), where the NG phase is stable in the absence of RSOC, we observe a phase transition to a homogeneous state with vanishing momentum (\( k = 0 \)) as the RSOC strength increases. This transition is reflected in the PMD, which initially remains flat for the NG phase but develops a peak at \( k = 0 \) upon introducing RSOC [Fig.~\ref{fig:plz}(a)]. As the RSOC strength increases, this peak becomes more pronounced. Additionally, the spin densities for both up-spin and down-spin fermions rise, indicating the onset of pairing facilitated by the RSOC, as shown in inset of Fig.~\ref{fig:plz}(a).
This pronounced spin imbalance, where the up-spin density ($n_{\uparrow}$) substantially trails the down-spin density ($n_{\downarrow}$), dictates an inevitable physical consequence: the system's structure is driven toward a homogeneous phase characterized by a zero momentum vector ($k=0$). This observation elegantly demonstrates how a fundamental asymmetry in density can effectively engineer the overall quantum ordering of the system.

To probe the topological properties of this homogeneous phase, we calculate the entanglement spectrum for a chain with 300 sites, as shown in Fig. 4(b). Our results reveal the appearance of entanglement degeneracy, a hallmark of topological phases. This degeneracy indicates the presence of nontrivial topology, even when the bulk energy spectrum remains gapped. Specifically, for \( \mu = 2 \) and \( h = 2 \), RSOC facilitates the transition from the NG phase to a homogeneous topological superfluid phase.

For \( h = 2 \) and \( \mu = 1 \), where the \emph{inter}-FFLO phase is stable without RSOC, we observe that the \emph{inter}-FFLO phase remains stable under weak RSOC strengths (\( \alpha \leq 0.15 \)), as illustrated Fig.~\ref{fig:plz}(c).
 As RSOC strength increases within the range \( 0.15 < \alpha < 0.5 \), the system transitions to a MP superfluid, characterized by both BCS and \emph{inter}-FFLO pairing. At higher RSOC strengths (\( \alpha \geq 0.5 \)), the MP superfluid phase eventually transitions entirely to a BCS phase. Our calculations of the entanglement spectrum, shown in Fig. 4(d), reveal that the \emph{inter}-FFLO phase remains topologically trivial as the RSOC strength increases, due to the absence of entanglement degeneracy.

The stability regions of the BCS, vacuum, and \emph{inter}-FFLO phases, as well as the transitions between them, are further explored in Fig. 5. As RSOC strength increases, the stability of the BCS phase expands, while the \emph{inter}-FFLO phase transitions to an MP phase and ultimately to the BCS phase at higher RSOC values.

Next, we extend our analysis to examine the role of DSOC in shaping the phase diagram [Fig.~\ref{fig:pt1}(a)], focusing on how the increase in DSOC strength influences the stability of the \emph{inter}-FFLO phase.
Our findings show that while the qualitative structure of the phase diagram remains largely unchanged with increasing DSOC, the stability of the \emph{inter}-FFLO phase is enhanced. The details of this analysis are shown in Fig.~\ref{fig:dsoc}. The introduction of DSOC strengthens the magnitude of the PMD in both BCS and \emph{inter}-FFLO phases while maintaining the overall topological features of the phase diagram.

The detailed dependence of fermion population \( n \), compressibility \( \chi_{\mu} \), spin densities \( n_{\uparrow}, n_{\downarrow} \), and spin polarization \( p \) on the chemical potential \( \mu \) for various DSOC strengths are presented in Figs.~\ref{fig:dsoc}(c-f). These results indicate that DSOC enhances the stability region of the \emph{inter}-FFLO phase, as indicated by the shift of anomalous peaks in compressibility to higher values of \( \mu \).
This behavior is expected, as equation (\ref{eq:kham}) demonstrates that DSOC functions as a momentum-dependent magnetic field, which induces spin polarization along the $z$-direction.

\begin{figure}[]
	\centering
	\includegraphics[scale=0.9]{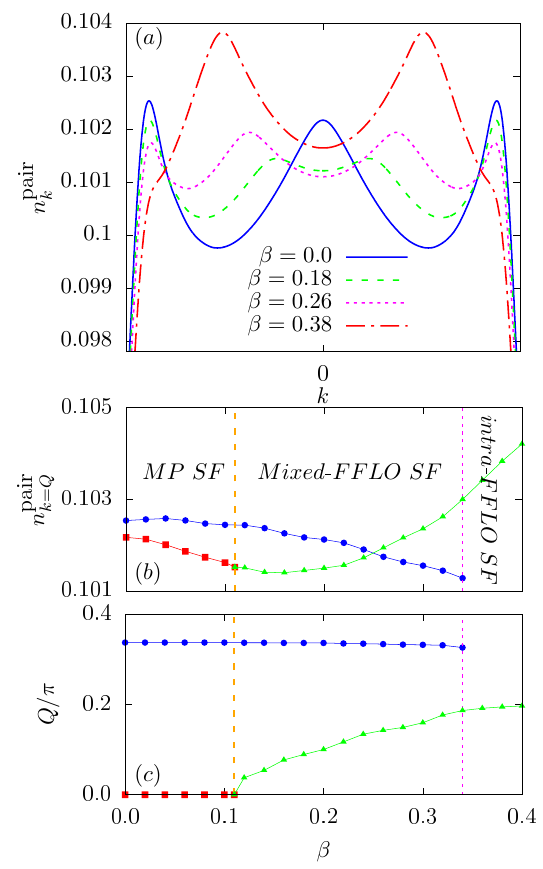}
	\caption{ (Color online)
		 (a) Pair Momentum Distribution, $n_{k}^{\text{pair}}$: Evaluation of the PMD in momentum space within the MP phase ($h = 2$, $\mu = 1$), as illustrated in the main phase diagram [Fig. 5(b)]. The influence of DSOC strength $\beta$ is depicted for $\beta = 0$ (solid line), $\beta = 0.18$ (dash-dotted line), $\beta = 0.26$ (dotted line), and $\beta = 0.36$ (dashed line). 
		(b) Peak PMD Amplitude: The maximum magnitude of $n_{k}^{\text{pair}}$ is plotted as a function of DSOC strength $\beta$. This maximum value is assessed at the momentum corresponding to the highest peak.
		(c) Evolution of the Peak Momentum: The center-of-mass momentum at which $n_{k}^{\text{pair}}$ reaches its maximum value, $k=Q$, is presented as a function of the DSOC strength $\beta$. Note that for small $\beta$, $Q=0$, consistent with the BCS phase, before shifting to finite momentum values indicative of FFLO states.
		}
	\label{fig:gm1}
\end{figure}
\subsection{Interplay Between RSOC and DSOC on the Phase Diagram of 1D Fermi Gases}

To gain a deeper understanding of the combined effects of transverse SOCs (RSOC and DSOC) on the stability and structure of superfluid phases, we investigate the phase diagram of 1D Fermi gases, employing the PMD, \( n^{\text{pair}}_k \). Specifically, we analyze how the PMD evolves within the MP phase as a function of DSOC strength (\(\beta\)), while keeping the Zeeman field (\(h = 2\)) and chemical potential (\(\mu = 1\)) fixed. These parameters correspond to a representative trajectory across the phase diagram shown in Fig. 5(b).

As illustrated in Fig.~\ref{fig:gm1}(a), at \(\beta = 0\), where only RSOC is present, the PMD exhibits a three-peak structure: a central peak at \(k = 0\), characteristic of intraband BCS-like pairing, and two symmetric peaks at finite momenta \(\pm k_{\text{FFLO}}\), indicative of \emph{inter}-FFLO correlations. This coexistence of pairing modes is a direct result of the band mixing induced by RSOC and the spin imbalance caused by the Zeeman field, defining the MP phase.

As the strength of DSOC, $\beta$, increases, several notable changes occur. The central peak in the PMD at $k=0$ diminishes rapidly, ultimately disappearing even at very low DSOC strengths. This phenomenon indicates a weakening of intraband BCS-like pairing, where fermions pair with zero center-of-mass momentum. In contrast to RSOC, DSOC tends to decouple the spin degrees of freedom, thereby destabilizing the BCS channel and promoting alternative pairing mechanisms.
At small DSOC strengths, the PMD displays nonuniform FFLO pairing, characterized by the presence of multiple or shifted peaks, which suggests a complex landscape of competing center-of-mass momenta. This behavior indicates a crossover from the MP phase to a \emph{mixed-FFLO} phase, accompanied by subtle band reconstruction. As the DSOC strength continues to increase, the finite momentum peaks in the PMD persist and intensify, confirming a transition to a state dominated by \emph{intra}-FFLO pairing. These pairings occur between fermions with momenta within the same helicity band, and their imbalance is further enhanced by the band shifting induced by DSOC.

To quantitatively characterize the phase transitions and the nature of pairing, we examine two key observables:

\begin{itemize}
	\item The peak amplitude of the PMD at the dominant pairing momentum $k=Q$, as shown in Fig.~\ref{fig:gm1}(b).
	\item The corresponding center-of-mass momentum $Q$, presented in Fig.~\ref{fig:gm1}(c), extracted from the maxima of $n_{k}^{\text{pair}}$. Note that $Q=0$ corresponds to the BCS phase at small $\beta$, while finite $Q$ indicates the presence of FFLO pairing.
\end{itemize}

The dependence of the dominant pairing momentum, $Q$, on the $\text{DSOC}$ strength $\beta$ in Fig.~\ref{fig:gm1}(c) reveals a crucial non-monotonic transition. At low $\beta$, $Q$ remains zero, confirming the stability of the $\text{BCS}$ pairing mechanism within the $\text{MP}$ phase. As $\beta$ increases, $Q$ progressively moves toward finite values, clearly signaling the emergence of finite-momentum pairing (the \emph{intra}-FFLO regime). This initial rise is consistent with theoretical expectations for weak DSOC, where the system transitions to a \emph{mixed}-FFLO state. Upon reaching higher $\text{DSOC}$ strengths ($\beta \approx 0.34$), the system enters a regime where the pure \emph{intra}-FFLO phase is stable. This indicates that DSOC acts as an effective quantum tuning parameter that drives a complete phase transition from the initial MP state to a purely \emph{intra}-FFLO state with a stable, finite center-of-mass momentum. This saturation behavior is consistent with theoretical expectations for strong DSOC [e.g., Figs.~\ref{fig:rd_spec}(e-f)].

\begin{figure}[]
	\centering
	\includegraphics[scale=.95]{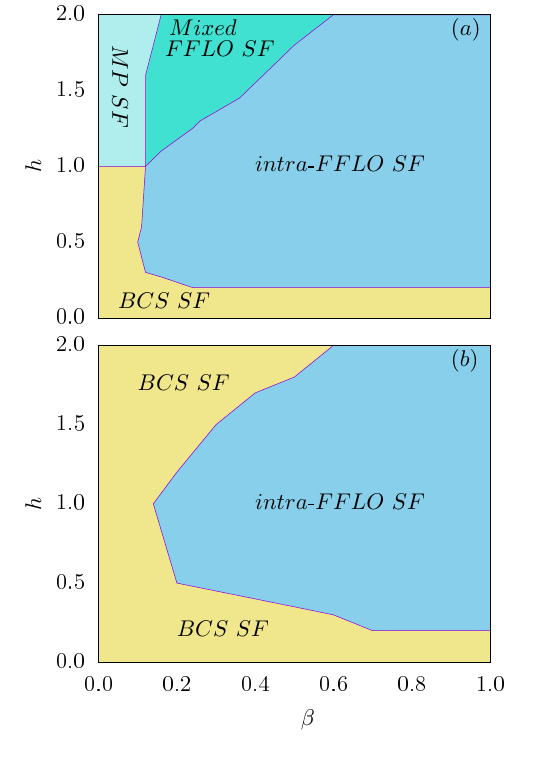}
	\caption{ (Color online)
		Phase diagrams of a 1D Fermi gas system in the $h-\beta$ plane. These diagrams are constructed for fixed parameters $U = -2$ and $\mu=1$, while varying the strengths of RSOC: (a) $\alpha=0.2$, and (c) $\alpha=0.6$.
		\label{fig:pdhb}		 }
	\label{fig:hy}
\end{figure}

The phase diagrams displayed in Fig.~\ref{fig:pdhb}, parametrized within the $h-\beta$ plane for RSOC strengths $\alpha = 0.2$ [Fig.~\ref{fig:pdhb}(a)] and $\alpha = 0.6$ [Fig.~\ref{fig:pdhb}(b)], illustrate the complex dependency of pairing phase stability and variety on the interplay between the Zeeman field $h$, RSOC, and DSOC $\beta$. Notably, the persistence of the BCS phase at low $h$, even under significant DSOC $\beta$, aligns with the characteristic behavior previously identified in Fig.~\ref{fig:rd_spec}(b).

In Fig.~\ref{fig:pdhb}(a), where $\alpha=0.2$, the system exhibits a rich set of finite-momentum superfluid phases at large \(h\): (i) MP with both zero- and finite-momentum components, (ii) a \emph{mixed-FFLO} phase with multiple finite center-of-mass momenta, and (iii) a pure \emph{intra}-FFLO phase characterized by two dominant peaks at \(\pm k_{\text{FFLO}}\).
 These phases arise due to the partial spin polarization induced by the Zeeman field and the momentum-selective nature of SOC, with DSOC increasingly favoring intraband pairing as it strengthens.

In contrast, for stronger RSOC (\(\alpha = 0.6\)) shown in Fig.~\ref{fig:pdhb}(b), the phase diagram simplifies significantly. The MP and \emph{mixed-FFLO} phases disappear entirely, leaving only two phases: the conventional BCS phase, and the pure \emph{intra}-FFLO phase.
This evolution highlights the contrasting roles of RSOC and DSOC: RSOC enhances intraband coherence, stabilizing zero-momentum BCS-type pairing while suppressing finite-momentum FFLO correlations. In contrast, DSOC reinforces spin polarization within helicity bands and energetically favors intraband pairing, thereby facilitating the emergence and stability of FFLO phases. The Zeeman field \(h\) further enhances spin imbalance, which is essential for FFLO pairing, but also suppresses BCS correlations when strong. The comparison between Fig.~\ref{fig:pdhb}(a) and \ref{fig:pdhb}(b) clearly shows that DSOC is crucial for stabilizing \emph{intra}-FFLO phases through its directional spin-locking mechanism.

The key insight from this analysis is that RSOC and DSOC have qualitatively different effects on pairing. RSOC promotes band mixing, allowing for hybrid pairing channels, while DSOC competes with RSOC and induces band decoupling. Their interplay results in a rich phase diagram not accessible in systems with only one type of SOCs. Specifically, DSOC acts as a control parameter that selectively enhances \emph{intra}-FFLO ordering while suppressing conventional superconductivity. This behavior could be exploited in engineered quantum materials or ultracold atomic platforms to control and manipulate quantum phase transitions.

 \begin{figure}[]
	\centering
	\includegraphics[scale=.95]{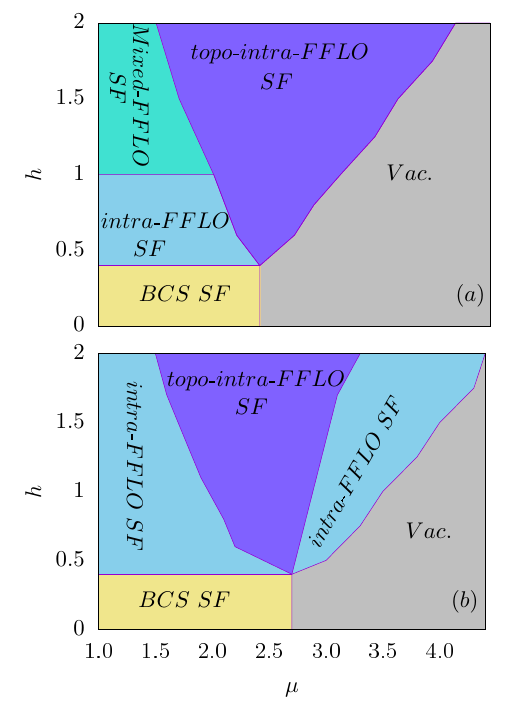}
	\caption{(Color online)
		Phase diagrams of a 1D Fermi gas system in the $h-\mu$ plane, with equal strengths for RSOC and DSOC and a fixed interaction strength of $U=-2$: (a) $\alpha=\beta=0.2$, and (b) $\alpha=\beta=0.6$.}
	\label{fig:9}
\end{figure}

Fig.~\ref{fig:9} presents the phase diagrams of a 1D Fermi gas system in the \( h-\mu \) plane, with equal strengths for RSOC and DSOC, and a fixed interaction strength \( U = -2 \). Panel (a), corresponding to weak RSOC and DSOC (\( \alpha = \beta = 0.2 \)), reveals that the system exhibits multiple phases, including the conventional BCS superfluid, the vacuum phase, the \emph{mixed-FFLO} superfluid, the trivial topological intraband FFLO (\emph{intra-FFLO}) superfluid, and a topologically nontrivial intraband FFLO (\emph{topo-intra-FFLO}) superfluid characterized by intraband finite-momentum pairings.
 As the Zeeman field \( h \) increases, the system transitions from the BCS phase to various FFLO phases, including the \emph{intra-FFLO} superfluid and the \emph{mixed-FFLO} superfluid. 
In comparison to Fig.~\ref{fig:pdhb}(b), where \( \alpha = 0.2 \) and \( \beta = 0 \), the MP phase vanishes, and the \emph{intra-FFLO} superfluid and \emph{mixed-FFLO} superfluid phases emerge at smaller values of the chemical potential (\( \mu \)).
Additionally, the stability region of the BCS phase decreases significantly with the introduction of DSOC, which plays a crucial role in stabilizing the FFLO superfluid phase. This result underscores the importance of DSOC in promoting finite-momentum pairing and spin imbalance, both of which are essential for the stability of the FFLO phase. The interplay between RSOC and DSOC in this regime results in a complex phase structure, where the homogeneous topological superfluid vanishes, and the \emph{topo-intra-FFLO} emerges by DSOC at intermediate values of $\mu$.

\begin{figure}
	\centering
	\includegraphics[scale=.95]{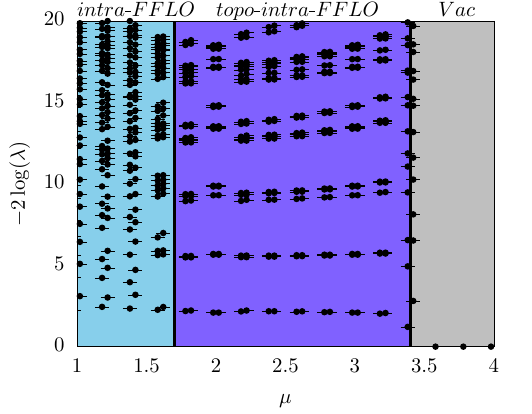}
	\caption{(Color online)
		Entanglement spectrum, $-2\log(\lambda)$, as a function of the chemical potential $\mu$ for a system with a fixed Zeeman field ($h=1.5$), based on the phase diagram shown in Fig.~\ref{fig:9}(a).}
	\label{fig:10}
\end{figure}

To explore the topological features of the induced-FFLO phase by DSOC, we calculate the entanglement spectrum for a chain of 300 sites, as shown in Fig.~\ref{fig:10}. The entanglement spectrum reveals the appearance of a twofold degeneracy, a clear signature of the topological nature of the intermediate phase, which we refer to as the \emph{topo-intra-FFLO} superfluid. This degeneracy confirms that the system, in this phase, exhibits robust topological protection.

The phase diagram shown in Fig.~\ref{fig:9}(b), computed for stronger RSOC and DSOC strengths ($\alpha = \beta = 0.6$), exhibits significant simplification. Under these conditions, the mixed-FFLO phase is suppressed, and the stability region of the pure \emph{intra}-FFLO phase expands into the high magnetic field regime. This stability results from the combined effect where strong DSOC induces spin polarization along the $z$-direction, and RSOC enhances intraband coherence, ultimately stabilizing the superfluid phase with vanishing center-of-mass momentum ($k=0$). Simultaneously, the stability region of the \emph{topo-intra}-FFLO phase in the $h-\mu$ plane contracts as the SOCs intensity increases. Notably, for $\mu \gtrsim 2.6$ and $h \gtrsim 0.5$, the \emph{topo-intra}-FFLO phase is replaced by the \emph{intra}-FFLO phase.

\section{Conclusion}

This study examines the significant role of RSOC and DSOC in stabilizing topological FFLO phases in a one-dimensional Fermi gas. Our findings indicate that DSOC is critical in stabilizing the \emph{topo-intra}-FFLO phase, a topologically nontrivial superfluid characterized by finite-momentum pairing within the same helicity band. While RSOC facilitates the formation of conventional zero-momentum Cooper pairs, DSOC uniquely enhances spin polarization and promotes the formation of FFLO states by suppressing interband coherence.

The presence of attractive interactions is vital for pairing, resulting in the emergence of various FFLO phases, including \emph{intra}-FFLO, \emph{mixed}-FFLO, and \emph{inter}-FFLO. The \emph{mixed}-FFLO phase, in particular, arises from the coexistence of both intraband and interband FFLO pairings, highlighting the intricate interplay between RSOC and DSOC.

Ultimately, our results contribute to a more comprehensive understanding of how the combined effects of RSOC, DSOC, and attractive interactions dictate the stability and emergence of these complex FFLO phases. These findings have substantial implications for the experimental realization of topologically nontrivial superfluids in ultracold atomic gases, thereby opening new avenues for their application in topological quantum computation. 

In summary, this work elucidates the mechanisms underlying the formation and stability of various FFLO phases, particularly the \emph{topo-intra}-FFLO phase, and lays the groundwork for the manipulation of topological phases in spin-orbit coupled systems.

\bibliography{References}

 \end{document}